\documentclass[times,twocolumn,final]{elsarticle}
\usepackage{colortbl}
\usepackage{adjustbox}
\usepackage{amsfonts}
\usepackage{amssymb,upref}
\usepackage{tgheros} 
\usepackage{tgtermes} 
\usepackage{anyfontsize}
\usepackage{enumitem}
\usepackage{subfig}
\usepackage{siunitx}

\usepackage{framed,multirow}
\usepackage{amssymb}
\usepackage{latexsym}
\usepackage{url}
\usepackage{xcolor}
\usepackage{hyperref}
\usepackage{mathrsfs}
\usepackage{graphicx}
\usepackage{subfig}
\usepackage{amsmath}
\usepackage{array}
\usepackage{booktabs}
\usepackage{threeparttable}
\usepackage{tabularx}
\usepackage[switch,columnwise]{lineno} 
\usepackage{rotating}
\usepackage{color} 
\usepackage{soul}
\usepackage{cite}
\usepackage{comment}
\soulregister\cite7 
\soulregister\ref7 
\soulregister\label7
\usepackage{balance}
\usepackage[utf8]{inputenc}
\usepackage[T1]{fontenc}

\definecolor{newcolor}{rgb}{.8,.349,.1}

\begin{document}
	\setlength{\parskip}{0.5em}
	\begin{frontmatter}
		\title{DeepNoise: Signal and Noise Disentanglement based on Classifying Fluorescent Microscopy Images via Deep Learning}
		\author[1]{Sen Yang \fnref{fn1}}
		\author[1]{Tao Shen \fnref{fn1}}
		\author[2]{Yuqi Fang \fnref{fn1}}
		\fntext[fn1]{Authors contributed equally to this work.}
		\author[3]{Xiyue Wang}
		\author[1]{Jun Zhang\fnref{fn2}}
		\author[1]{Wei Yang}
		\author[1]{Junzhou Huang}
		\author[1]{Xiao Han \fnref{fn2}}
		\fntext[fn2]{Corresponding authors: Jun Zhang (xdzhangjun@gmail.com) and Xiao Han (haroldhan@tencent.com)}
		
		\address[1]{Tencent AI Lab, Shenzhen 518057, China}
		\address[2] {Department of Electronic Engineering, The Chinese University of Hong Kong, Shatin, New Territories, 999077, Hong Kong}
		\address[3]{College of Computer Science, Sichuan University, Chengdu 610065, China}

\begin{abstract}


The high-content image-based assay is commonly leveraged for identifying the phenotypic impact of genetic perturbations in biology field. However, a persistent issue remains unsolved during experiments: the interferential technical noise caused by systematic errors (e.g., temperature, reagent concentration, and well location) is always mixed up with the real biological signals, leading to misinterpretation of any conclusion drawn. Here, we show a mean teacher based deep learning model (DeepNoise) that can disentangle biological signals from the experimental noise. Specifically, we aim to classify the phenotypic impact of 1,108 different genetic perturbations screened from 125,510 fluorescent microscopy images, which are totally unrecognizable by human eye. We validate our model by participating in the Recursion Cellular Image Classification Challenge, and our proposed method achieves an extremely high classification score (Acc: 99.596\%), ranking the 2nd place among 866 participating groups. This promising result indicates the successful separation of biological and technical factors, which might help decrease the cost of treatment development and expedite the drug discovery process.

\end{abstract}


	\end{frontmatter}

\section{Introduction}

Usually, it takes over 10 years and billions of dollars to find a new drug. Gene knockout, a genetic technique which can make an organism's genes inoperative, is widely used as a screening tool in accelerating drug discovery and development.
Gene knockout based on RNA interference (RNAi) can be achieved by introducing small interfering RNA (siRNA), 
which is designed to be fully complementary to a portion of an mRNA, interfering with the gene and protein expression \citep{elbashir2001duplexes}.
SiRNA transfection is to perturb the morphology, such as count of cells, creating a distinctive phenotype corresponding to each siRNA. To further investigate the results of siRNA transfection, high-content imaging techniques are often used to screen, visualize and quantitatively analyze cellular feature representations \citep{conrad2010automated, boutros2015microscopy}, which have been widely adopted in many field of biology, e.g., genetics \citep{zhou2014high, echeverri2006high}, drug discovery and development \citep{swinney2011were, broach1996high, boutros2015microscopy, macarron2011impact}. In this study, we adopt \textit{Cell Painting} \citep{bray2016cell}, a high-content image-based assay for morphological profiling, to identify the phenotypic impact of genetic perturbations. Specifically, cells perturbed with different treatments are plated in multi-well plates, which are imaged on a high-throughput microscope using multiplexed fluorescent dyes. By analyzing these images, morphologically relevant similarities and differences among samples caused by genetic perturbations can be identified, and the valuable biological information about cellular state can be captured.

\begin{figure*}[!t]
	\includegraphics[width=\textwidth]{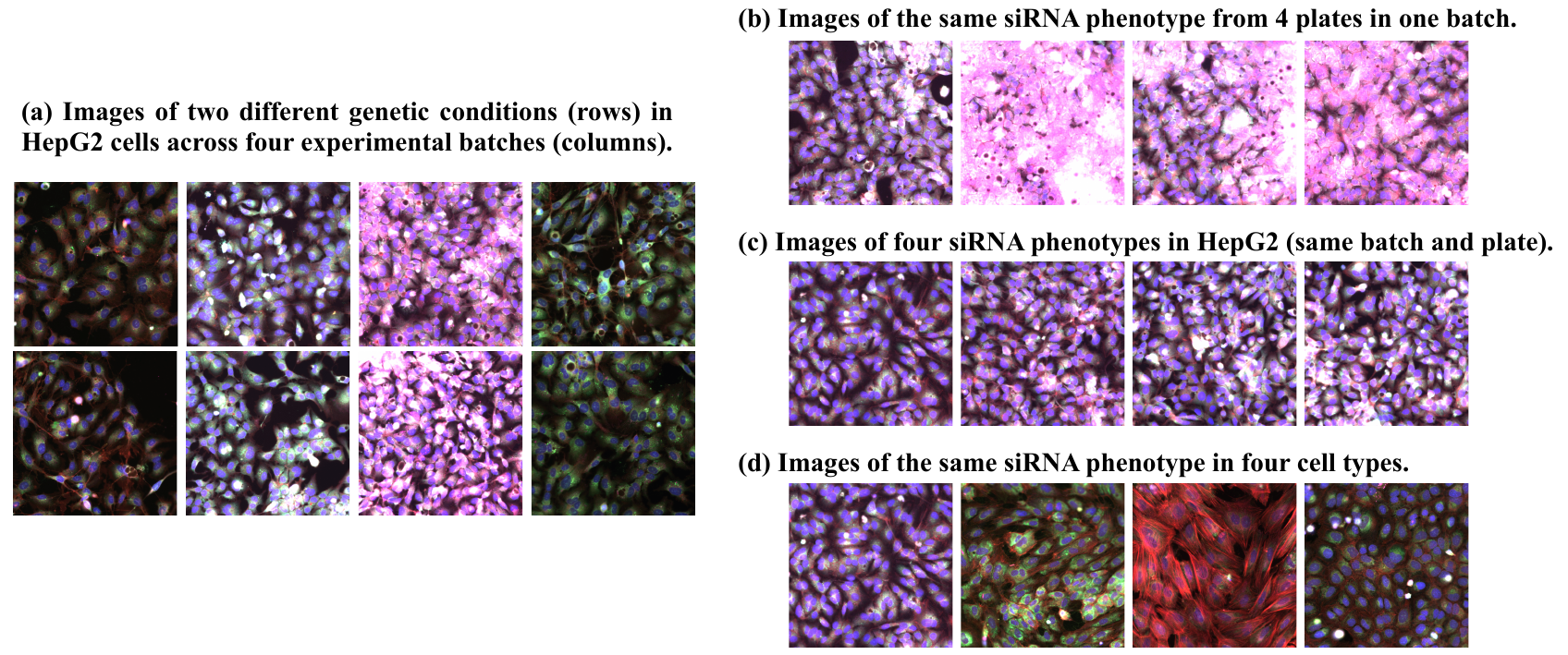}
	\centering
	\caption{The human have much difficulty in identifying the phenotypic impact of genetic perturbations. (a) The demonstration of microscopy images generated from two different siRNAs (rows) in HepG2 cells across four different batches (columns). Four morphological phenotypes in each row are all derived from the same siRNA targeting the same cell but in different experimental batches. There exists large visual difference across four batches (i.e, batch effects). Different siRNAs may generate visually similar expressions by comparing morphological phenotypes in each column, which introduces large disturbance on real biological signal capture. (b) The demonstration of microscopy images generated from the same siRNA for four plates in one experimental batch. Four phenotypes are all derived from the same siRNA targeting the same cell in one batch but in different plates. Visually different phenotypes are shown across four plates (i.e., plate effects). (c) The demonstration of phenotypes generated from four distinct siRNAs from the same experiment and the same plate, which are totally unrecognizable by human eye. (d) The demonstration of microscopy images generated from the same siRNA in four different cell types.}
	\label{fig_batchplate_effects}
\end{figure*}


However, there exists a big challenge in identifying cellular phenotypes monitored in the readout assay, i.e., technical noise (batch effects and plate effects) can significantly invalidate the biological conclusion drawn. For batch effects, even the experiments are carefully designed to control for systematic variables (e.g., temperature, humidity), the biological measurements derived from the assay screens can mix up with the non-biological artifacts (see Fig.~\ref{fig_batchplate_effects}(a)). For plate effects, phenotypes are distinct across different plates even they are derived from the same siRNA targeting the same cell in the same batch (Fig.~\ref{fig_batchplate_effects}(b)). Both batch effects and plate effects are inherent and unavoidable during the execution of biological experiments. Additionally, as demonstrated in Fig.~\ref{fig_batchplate_effects}(c)-(d), it is impossible to detect phenotypes generated from the same/distinct siRNAs by human eye.

In order to capture real biological signals from measurements taken from high-throughput screens, effectively eliminating these noises is in highly demand \citep{soneson2014batch, leek2010tackling, nygaard2016methods, parker2012practical}. \textit{R$\times$R$\times$1} \footnote{https://www.rxrx.ai/rxrx1} is the first publicly-available biological dataset that is systematically created to study this noise-removal problem. Our study adopts this dataset and tackles the task that classifies 125,510 screened microscopy images of cells under one of 1,108 different genetic perturbations.
A high classification score suggests the biological and technical factors can be effectively separated.

\begin{figure*}[!t]
	\includegraphics[width=\textwidth]{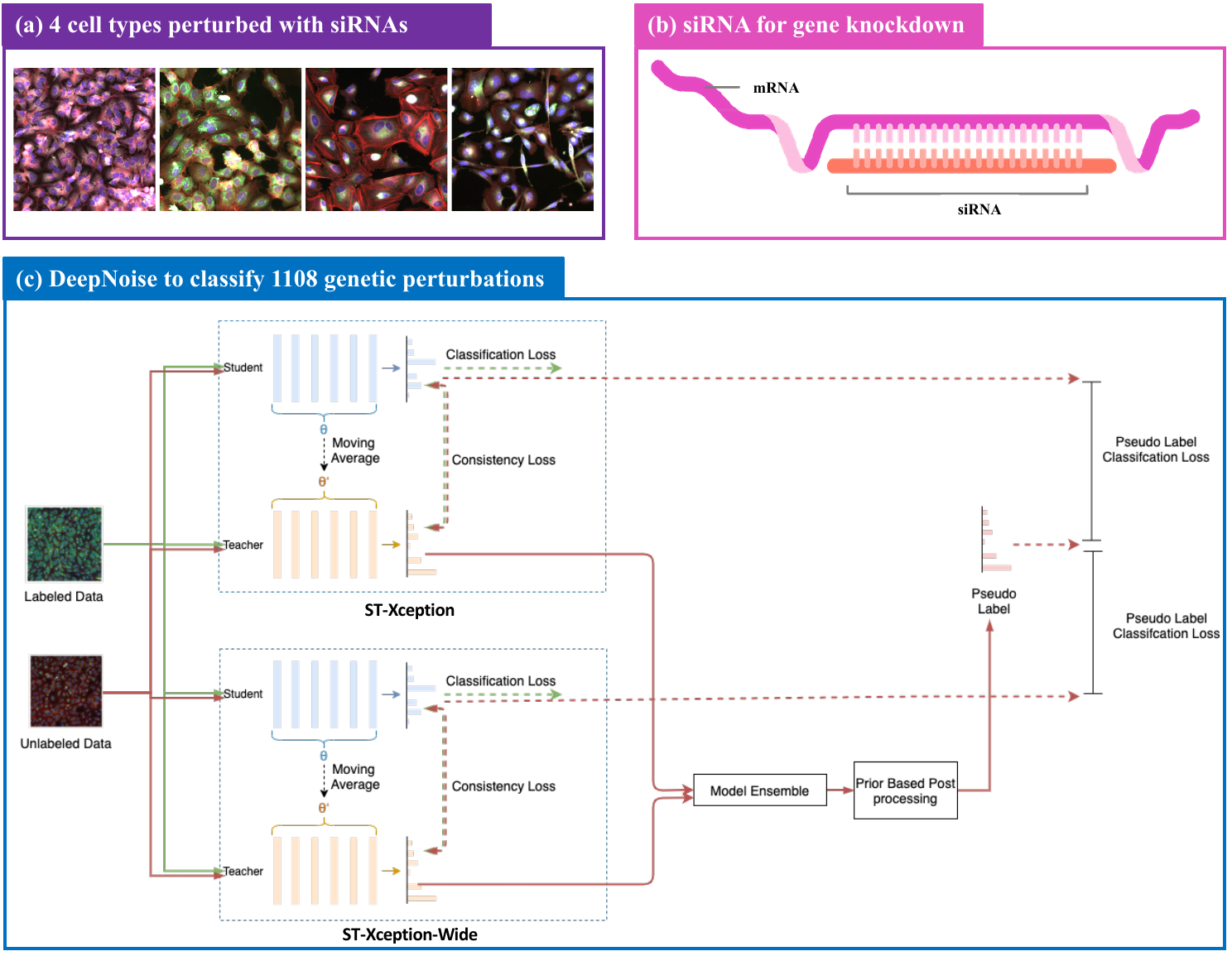}
	\centering
	\caption{The full pipeline of this study. (a) Fluorescent microscopy images generated from four distinct cell types perturbed with different siRNAs, showing much diverse phenotypes. (b) A demonstration of full-complimentarity of an siRNA to an mRNA to knockdown a particular target gene. (c) The proposed DeepNoise network architecture, which is based on mean teacher strategy \citep{tarvainen2017mean}, a semi-supervised teacher-student deep learning network that averages the model weights of the student network. The classification loss, i.e., $L_{ArcFace}$, is applied on the student network, and a consistency loss, i.e., $L_{Consist}$, is used to minimize the difference between the outputs of the student and the teacher network. Two models (Xception and Xception-Wide) are integrated and the final classification result is derived by averaging these two models' prediction outputs. The detailed description of the model can refer to ``DeepNoise in 1,108 genetic perturbation classification.'' section in Results.}
	\label{fig_flowchart}
\end{figure*}

To achieve that, we in this work develop an intelligent deep learning based model, i.e., \textit{DeepNoise}. Traditional methods of classification tasks mainly depend on the handcrafted features, e.g., shapes \citep{kothari2013histological, zhang2004review, krishnan2012computer} and textures \citep{nanni2012survey, desir2012svm}. Whereas the acquisition and quantification of these features highly depend on domain knowledge and manual design, the accuracy and robustness of traditional methods remain unsatisfying. For instance, phenotypes of images perturbed with two different siRNAs are visually similar, thus it becomes quite difficult for traditional approaches to extract the distinct features for each siRNA (see Fig.~\ref{fig_batchplate_effects}(a)). Recently, regarding the prosperity of deep learning in automatic feature mining, many studies have applied this kind of artificial intelligent technique for image classification \citep{bayramoglu2016deep, coudray2018classification, korbar2017deep, hou2016patch}. Compared with the handcrafted ones, features learned by deep learning methods are with much diversity and may mine for the inherent difference in the phenotypic profiles induced by different siRNAs, enabling a better generalization ability of the models. The proposed deep learning model is presented in Fig.~\ref{fig_flowchart}(c), and it achieves a classification accuracy of 99.596\% in this challenging 1,108 classification task. The near-perfect result indicates the effectiveness of our proposed method on disentangling biological signals from the experimental noise, which may dramatically decrease the cost of treatment development and expedite the process of drug discovery. 

\section*{Results}\label{Results}
\textbf{DeepNoise in 1,108 genetic perturbation classification.} In this study, we propose a semi-supervised deep learning network (DeepNoise) to identify the phenotypic impact of 1,108 genetic perturbations. The network architecture is comprised of two base models (see Fig.~\ref{fig_flowchart}(c)), i.e., student-teacher-Xception (ST-Xception) and ST-Xception-Wide (see ``Model ensemble'' section in Methods), which are both based on mean-teacher strategy \citep{tarvainen2017mean}. Taking ST-Xception (top of Fig.~\ref{fig_flowchart}(c)) for example, the student network adopts Xception module \citep{chollet2017xception}, which replaces Inception modules with depthwise separable convolutions, enriching the diversity of learned feature representations. The teacher network sharing the same architecture with the student network is initialized based on pretrained weights of ImageNet \citep{deng2009imagenet}. 
During model training, the model weights of student network are first averaged and passed to the teacher network, and then the teacher network is updated by combining the updated student network with the historical information of teacher network using an exponential moving average strategy.
The student network learns from the teacher network by minimizing the classification loss computed from the input annotated data (i.e., $L_{ArcFace}$, see ``Arcface loss'' section in Methods), and the consistency loss between the teacher and student networks computed from the unannotated data (i.e., $L_{Consist}$, see ``Consistency loss'' section in Methods). 

For ST-Xception-Wide (bottom of Fig.~\ref{fig_flowchart}(c)), both student and teacher network use Xception-Wide, which has wider convolutional channels and larger model capability compared with Xception \citep{chollet2017xception}. To achieve the final prediction, the results of both ST-Xception and ST-Xception-Wide are averaged, and go through a prior based post processing step (see ``Prior based post processing'' section in Methods) to derive the pseudo labels. The pseudo labels update using the highest validation score generated by teacher networks during the training procedure. Eventually, we introduce $L_{pSoftmax}$ (see ``Pseudo label softmax loss'' section in Methods) to minimize the difference between the averaged prediction derived from two teacher networks and the output of each student network, which can be regarded as another consistency loss function.

\begin{figure*}[!t]
	\includegraphics[width=\textwidth]{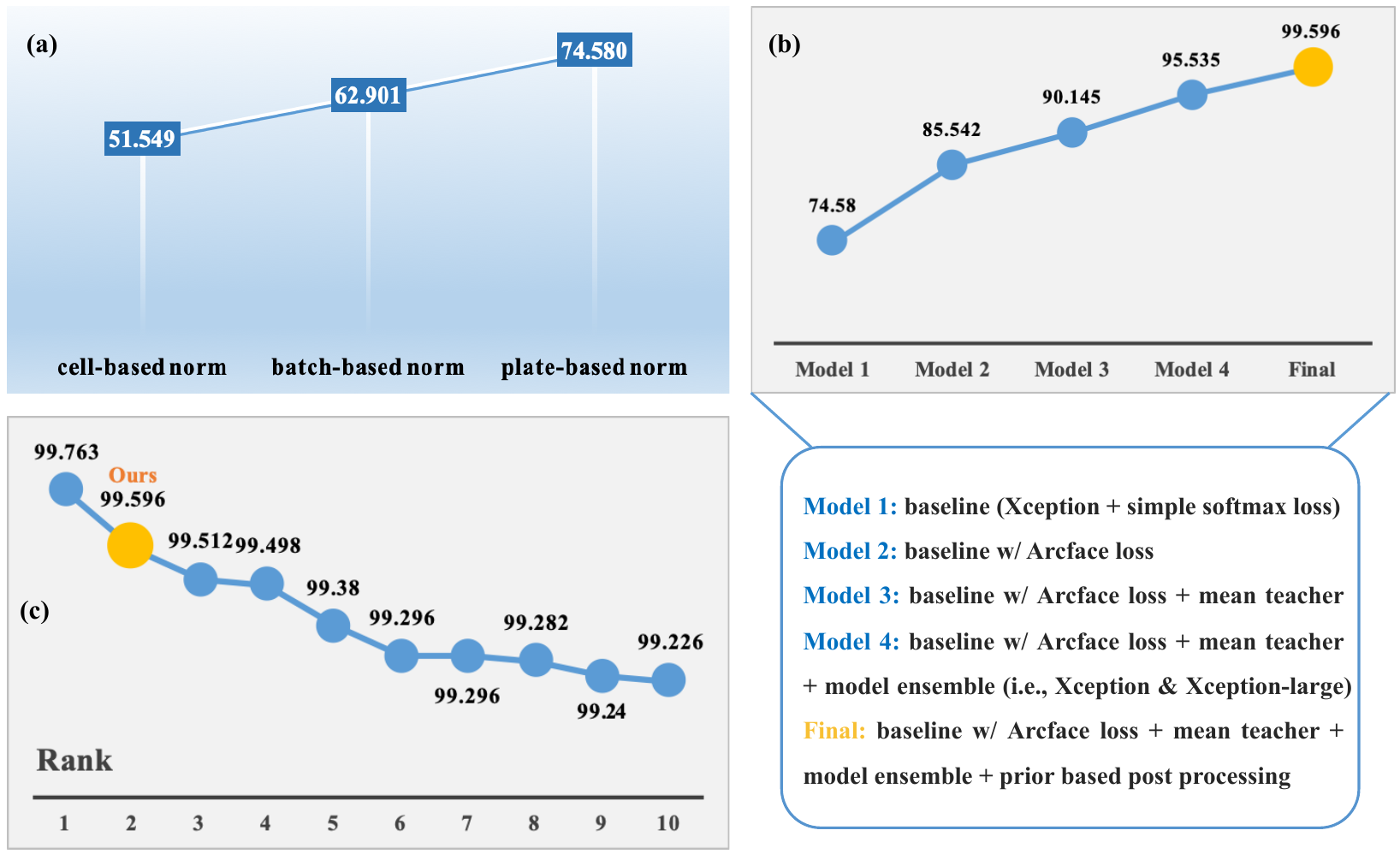}
	\centering
	\caption{(a) Prediction results under different normalization strategies in pre-processing step. (b) Ablation studies to demonstrate each component introduced into DeepNoise. The detailed model description can refer to Fig.~\ref{fig_ablationstudy}. (c) Leaderboard of \textit{CellSignal} competition. Our proposed DeepNoise ranked 2nd place among 866 participants with a classification accuracy of 99.596\%.}
	\label{fig_acc_norm_ablat_lebd}
\end{figure*}

\textbf{Different normalization strategies in pre-processing step.} The adopted dataset is comprised of 125,510 fluorescence microscopy images of human cells of four different types, i.e., HUVEC, RPE, HepG2, and U2OS, and it is collected from 51 batches with 4 plates in one batch. Since phenotype profiles produced visually differ from each other, it becomes necessary to apply normalization on these microscopy images to mitigate this variation to some extent. In this work, we explore three normalization strategies before model training, i.e., cell-based normalization, batch-based normalization, and plate-based normalization, and the model prediction results under three strategies are shown in Fig.~\ref{fig_acc_norm_ablat_lebd}(a). Except the difference of normalization strategy, all training models are the same, i.e., Xception \citep{chollet2017xception} optimized with the simple softmax loss function.

\textit{Cell-based normalization.} Cell-based normalization is to calculate the mean value and standard deviation (std) for microscopy images of each cell type, and then standardize these images based on their corresponding cell type (see Eq.~\ref{result_cellbased_norm}).

\begin{equation}\label{result_cellbased_norm}
x_{cell\_norm} = \frac{x - mean_i}{std_i}
\end{equation}
where $i$ means HUVEC, RPE, HepG2, and U2OS. $x$ represents the input image; $mean_i$ and $std_i$ denotes the mean and std of the corresponding cell type of the input image.

\textit{Batch-based normalization.} There are totally 51 experimental batches in this study, and batch effects inevitably create factors of variation within the data that are irrelevant to the biological information, even when these batches are carefully designed to control for technical variables. Therefore, we consider conducting the batch-based normalization for these microscopy images (see Eq.~\ref{result_batchbased_norm}).

\begin{equation}\label{result_batchbased_norm}
x_{batch\_norm} = \frac{x - mean_j}{std_j}
\end{equation}
where $j$ means one of 51 experimental batches; $mean_j$ and $std_j$ denotes the mean and std of the corresponding batch of the input image.

\textit{Plate-based normalization.} Since plate effect also interferes with the biological conclusion drawn, we apply the third strategy, i.e., plate-based normalization, on these microscopy images (see Eq.~\ref{result_platebased_norm}).

\begin{equation}\label{result_platebased_norm}
x_{plate\_norm} = \frac{x - mean_k}{std_k}
\end{equation}
where $k$ means one of 204 (51$\times$4) experimental plates; $mean_k$ and $std_k$ denotes the mean and std of the corresponding plate of the input image.

The prediction results of 1,108 genetic perturbations based on these three normalization methods are displayed in Fig.~\ref{fig_acc_norm_ablat_lebd}(a). We observe that only 51.549\% classification accuracy is achieved under the cell-based normalization. This result is expected because neither patch effects nor plate effects are considered to be removed, thus biological information and interferential technical noise in experiments are mixed up. Batch-based normalization is developed for tackling batch effects, using which the classification accuracy reaches 62.901\%. Although there exists a dramatic improvement of the prediction accuracy compared with cell-based normalization method, this batch-based normalization strategy may not be optimal because four plates in one batch still affect each other. To tackle this problem, we finally apply plate-based normalization on this dataset, and the prediction accuracy of 1,108 genetic perturbations reaches up to 74.580\%.


\begin{figure*}[!t]
	\includegraphics[width=\textwidth]{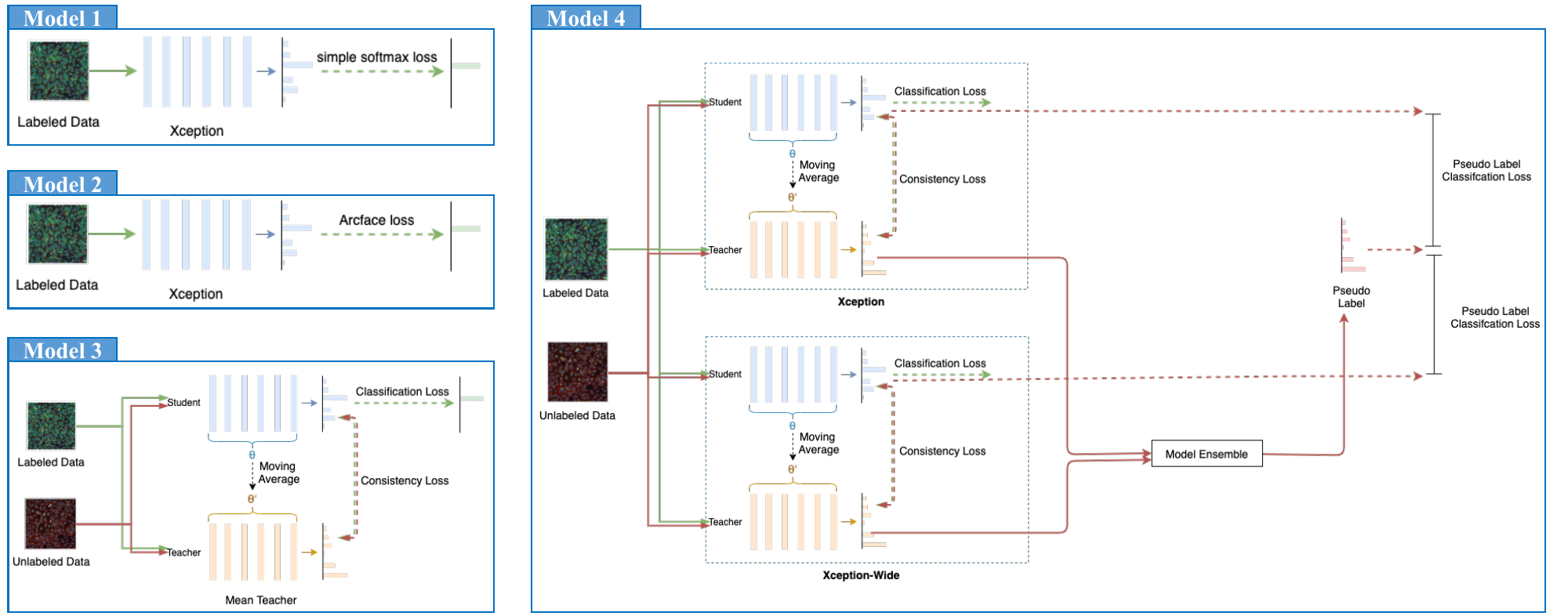}
	\centering
	\caption{Four ablation studies to demonstrate each component introduced into DeepNoise. Model 1: Xception optimized with a simple softmax loss function, which is regarded as a baseline. Model 2: Replacing the simple softmax loss in Model 1 with the arcface loss. Both Model 1 and Model 2 are fully-supervised models. Model 3: A semi-supervised mean-teacher strategy is utilized, in which both the student and the teacher network adopt Xception. For annotated data (training microscopy images), a classification loss, i.e., $L_{ArcFace}$, and a consistency loss, i.e., $L_{Consist}$, are jointly to train the network, while for unannotated data (testing microscopy images), only the consistency loss, i.e., $L_{Consist}$, is adopted to optimize the network. Model 4: a model ensembling strategy is leveraged to aggregate predictions of two base models, i.e, ST-Xception and ST-Xception-Wide. The pseudo labels update using the highest validation score generated by two teacher networks during the training procedure. The pseudo label classification loss function (see ``Pseudo label softmax loss'' section in Methods) is leveraged to minimize the difference between the averaged predictions derived from two teacher networks and the output of each student network.}
	\label{fig_ablationstudy}
\end{figure*}

\textbf{Ablation studies.} We report each component introduced into DeepNoise network that improves the prediction accuracy of genetic perturbation classification. The following ablation studies are all based on plate-based normalization strategy (see ``Different normalization strategies in pre-processing step'' section in Results). The baseline model (Model 1 in Fig.~\ref{fig_ablationstudy}) is Xception \citep{chollet2017xception} optimized with a simple softmax loss function, which achieves a classification accuracy of 74.580\%. For Model 2, we replace the simple softmax loss in Model 1 with the arcface loss \citep{deng2019arcface}. Since the number of classes we identify is quite large (i.e., 1,108 classes), how to enhance the intra-class compactness and inter-class discrepancy becomes important. The arcface loss \citep{deng2019arcface} added with an angular margin is really good at separating inter-class distance, thus it is leveraged to optimize the neural network. Compared with Model 2 and Model 1 (see Fig.~\ref{fig_acc_norm_ablat_lebd}(b)), we can see that an accuracy of 14.7\% ((85.542-74.58)/74.58$\times$100\%) is improved when using arcface loss. For both Model 1 and 2, any information of the unannotated (test) fluorescent microscopy images is not taken in account, that is, both models are fully-supervised.

The following three models are the semi-supervised approaches, which consider utilizing the unannotated (test) microscopy images to assist the model training. Model 3 (see Fig.~\ref{fig_ablationstudy}) is designed based on mean teacher strategy \citep{tarvainen2017mean}. Both student and teacher network adopt Xception \citep{chollet2017xception} as the training network. For annotated data (training microscopy images), a classification loss and a consistency loss are jointly to train the network, while for unannotated data (testing microscopy images), only the consistency loss is adopted to optimize the network. Comparing the classification accuracy derived from Model 2 and 3 (see Fig.~\ref{fig_acc_norm_ablat_lebd}(b)), we can see that feature representations derived from unannotated images indeed enhance the prediction accuracy. To further improve the classification performance, we adopt the model ensembling strategy by aggregating predictions of two base models, i.e, ST-Xception and ST-Xception-Wide (Model 4 in Fig.~\ref{fig_ablationstudy}). By model ensemble, the strengths of both models are taken advantage of, and the features generated from two teacher networks can be more representative. By aggregating the complementary information of two models, the classification accuracy can be significantly improved (from 90.145\% to 95.535\%, see Fig.~\ref{fig_acc_norm_ablat_lebd}(b)). Eventually, we apply a prior based post processing (see ``Prior based post processing'' section in Methods) before pseudo label generation (see Fig.~\ref{fig_flowchart}(c)), which takes into consideration the prior information that same siRNA does not occur twice in an experimental plate. As expected, the classification accuracy increases after all classes are equally balanced (from 95.535\% to 99.596\%, see Fig.~\ref{fig_acc_norm_ablat_lebd}(b)). 

\begin{figure*}[!t]
	\includegraphics[width=\textwidth]{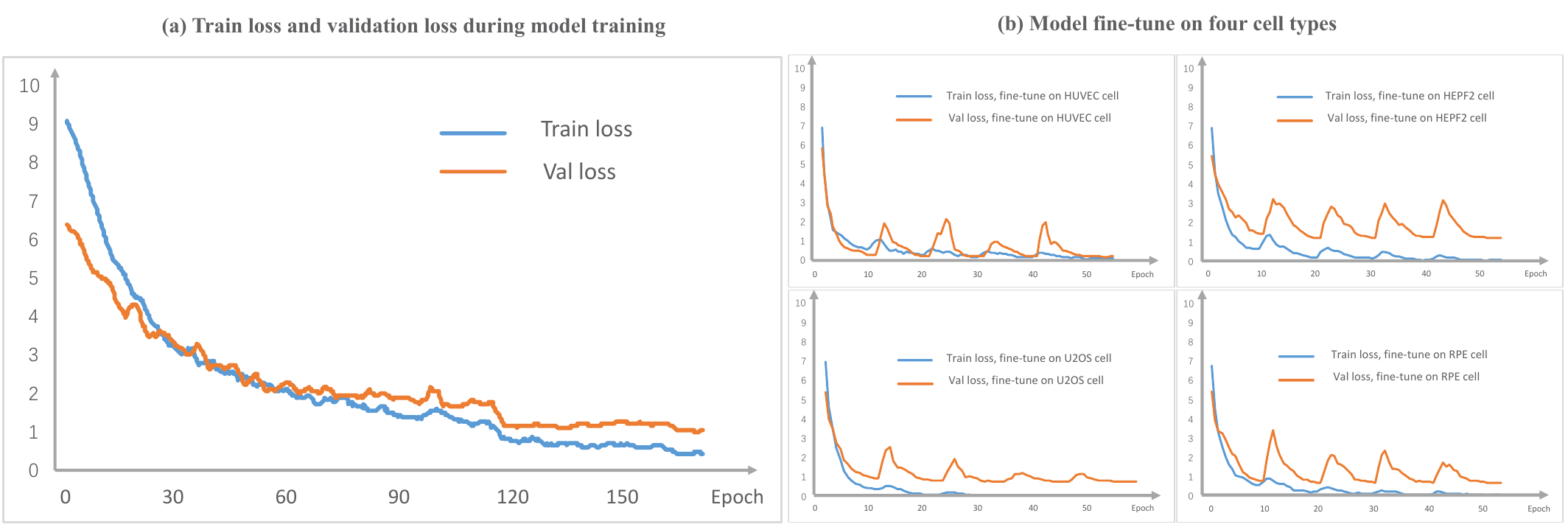}
	\centering
	\caption{The training and validation loss curves during training process. (a) All microscopy images no matter what the cell type they belong to are used for training. (b) The training loss and validation curves fine-tuned on four cell types respectively. The fine-tuning is conducted after 160 training epochs under (a).} \label{fig_trainval_loss_curve}
\end{figure*}

The training and validation loss curves during training process are displayed in Fig.~\ref{fig_trainval_loss_curve}. At the beginning, all microscopy images no matter what the cell type they belong to are used for training, and the training loss and validation loss are shown in Fig.~\ref{fig_trainval_loss_curve}(a). With the aim of extracting the cell-specific feature representations, we fine-tune DeepNoise on four cell types respectively. Fig.~\ref{fig_trainval_loss_curve}(b) shows the training loss and validation loss in terms of four cell-specific models. During the inference procedure, for the microscopy images under the specific cell types, we apply the corresponding trained model and generate the final prediction results.

\textbf{Comparison with top ten teams participating in the challenge.} The organizer of R$\times$R$\times$1 dataset sponsored a challenge named \textit{CellSignal} to encourage researchers to explore methods of disentangling biological signals from technical noise. Totally 866 teams participated in this competition, and the scores of top ten teams of the leaderboard is shown in Fig.~\ref{fig_acc_norm_ablat_lebd}(c). Our proposed DeepNoise ranked 2nd place, achieving a classification accuracy of 99.596\%. According to the solution of the competition champion released on the official website\footnote{https://www.kaggle.com/c/recursion-cellular-image-classification/discussion/110543}, we find that they adopted an ensemble of 11 deep learning models (6$\times$DenseNet-161 and 5$\times$DenseNet-201 \citep{huang2017densely}), while our proposed solution only integrated two models (ST-Xception and ST-Xception-Wide, see ``Model ensemble'' section in Methods) to derive the final classification result. It is likely that the increasing number of ensemble models contributes to their championship.

\section*{Discussions}\label{Discussions}
In this work, we develop a deep learning based model (DeepNoise) that can disentangle the real biological signals from the interferential technical noise by identifying phenotypic impact of 1,108 different genetic perturbations screened from 125,510 fluorescent microscopy images. The proposed method achieves an extremely high classification score, with a multi-class accuracy of 99.596\%. Moreover, among 866 participating groups that adopt the same database, our method competes favorably and wins the second prize. This promising result verifies the successful separation of biological and technical factors, helping the biology researchers derive conclusions from real biological information. Moreover, the persistent issue that biological experiments interfered with batch effects are hardly reproducible is also tackled to some extent.

Our proposed DeepNoise is a deep learning based model. Compared with conventional methods that depend on handcrafted features for classification (shape, color, and texture), deep learning methods can automatically discover intricate hierarchical feature representations for a large quantity of input data. Deep learning technique has been adopted in many biological fields, including gene expression modeling \citep{chen2016learning, chen2016gene}, protein structure prediction \citep{zhou2015predicting}, DNA methylation \citep{wang2016predicting}, and protein localization \citep{parnamaa2017accurate}. Our proposed model is based on the semi-supervised mean-teacher strategy \citep{tarvainen2017mean}, which has achieved the state-of-the-art performance on both CIFAR-10 and ImageNet 2012 challenges. Thanks to the averaged weight from student to teacher network, mean-teacher strategy can effectively exploit the unlabeled data, which exactly suits our needs. In future work, we would like to explore other semi-supervised \citep{berthelot2019mixmatch, odena2016semi} or self-supervised learning models \citep{zhai2019s4l, noroozi2018boosting} to tackle this genetic perturbation problem.


The limitations of this study are two aspects. Firstly, since the labels of test dataset are not accessible, extensive comparisons between our proposed DeepNoise and other classification models cannot be conducted. The more detailed investigations also cannot be explored, including the predictions in each cell type, other measurement metrics (e.g., specificity, sensitivity, precision) that further validate the proposed method. Secondly, our submitted solution does not consider any information of 30 control genes (see ``Datasets'' section in Methods) in each experimental plate, which are totally the same across different plates. This information is believed to be valuable. Our future work will seek to intelligently combine features derived from these reference images and current learned features to further improve the classification accuracy. 

\section*{Methods}\label{Methods}
\textbf{Datasets.} In this study, we adopt the R$\times$R$\times$1 dataset\footnote{https://www.rxrx.ai/rxrx1} released by \textit{Recursion Pharmaceuticals}\footnote{https://www.recursionpharma.com/}. This dataset contains 125,510 fluorescent microscopy images representing 1,108 classes. Each fluorescent microscopy image contains 6 channels with different wavelengths, which respectively represent 6 different cell organelles, i.e., the nucleus, nucleolus, mitochondria, endoplasmic reticulum, actin cytoskeleton, and golgi apparatus \citep{bray2016cell}.

The entire dataset consists of 51 experiments, and each experiment is executed in one batch. Each batch holds a single cell type: 24 in HUVEC, 11 in RPE, 11 in HepG2, and 5 in U2OS. There are four plates in one batch, each with 384 (16$\times$24) wells. One of 1,108 different siRNAs is introduced into each well to create distinct genetic conditions. Images located in the outer rows and columns of the plate are not utilized since they are hugely affected with environmental effects, hence for each plate there remain 308 wells. Each plate contains the same 30 control siRNAs, 277 different perturbed siRNA, and one intact well. In each experiment, the locations of 1,108 (277$\times$4) perturbed siRNAs are randomized. \textit{Recursion Pharmaceuticals} releases this dataset and held a competition on Kaggle\footnote{https://www.kaggle.com/c/recursion-cellular-image-classification}, attracting 866 teams from all over the world to participate in.

\textbf{Evaluation metric.} As the challenge organizers specified, \textit{multi-class accuracy (Acc)} is adopted to evaluate the model performance, which is simply the average number of observations with the correct label. The metric improves with the increased number of correctly classified images.

\textbf{Training details.} For model training, a mini-batch of size 64 is adopted and the Adam optimizer \citep{kingma2014adam} with weight decay (2e-4) is used as the optimization method. The initial learning rate is set to 3e-4, which is reduced to 1e-4 after 100 training epochs. After 140th epoch the learning rate decreases by a factor of 10, and there are totally 160 training epochs. All models are performed using the PyTorch package \citep{paszke2017automatic} and all experiments are implemented on a workstation equipped with four 24 GB memory NVIDIA Tesla P40 GPU cards.

Standard real-time data augmentation methods such as horizontal flip, vertical flip, $90^{\circ}$ flip, random erasing, and random scale are performed to make the model learn features invariant to geometric perturbations. 

\textbf{Arcface loss.} The arcface loss $L_{ArcFace}$ \citep{deng2019arcface} is leveraged to minimize the difference between prediction outputs of the student network and the one-hot classification label. Compared with other classification cost functions, e.g., softmax loss and its variants \citep{wen2016discriminative, deng2017marginal, zhang2017range, wang2018additive, liu2017sphereface, wang2018cosface}, the arcface loss we adopt incorporates an additive angular margin in the loss function (see Eq.~\ref{eq_arcface}) to enforce extra intra-class compactness and inter-class discrepancy simultaneously.

\begin{equation}\label{eq_arcface}
L_{ArcFace}= -\frac{1}{N}\sum_{i=1}^N log\frac{e^{s(cos(\theta_{y_i}+m))}}{e^{s(cos(\theta_{y_i}+m))} + \sum_{j=1,j \neq y_{i}}^n e^{s\cos\theta_{j}}}
\end{equation}
where $i$-th sample, belonging to the $y_{i}$-th class; $N$ is the number of classes. $\theta_{y_i}$ is the angel between the weight $W_{j}$ and the feature $x_{i}$, herein, $\left \| W_{y_i} \right\|$ is fixed to 1, and the embedding feature $\left \| x_{i} \right\|$ is normalized and re-scaled to $s$ ($s = \left \| x_{i} \right\|$). $m$ is an additive angular margin between $x_{i}$ and $W_{y_i}$ to enforce intra-class compactness and inter-class discrepancy simultaneously. \textit{s} and \textit{m} are hyper-parameters. Experimentally, \textit{s} is set to 30, and \textit{m} is set to 0.1.

\textbf{Consistency loss.} The consistency loss $L_{Consist}$ is designed to minimize the difference between the prediction of the student network (with weights $\theta$) and that of the teacher network (with weights $\theta^{'}$), which is defined as follows:

\begin{equation}\label{eq_consisloss}
L_{Consist} = \mathbb{D}_{x} \Big[ \big \| f(x, \theta^{'})-f(x, \theta) \big \|^{2} \Big]
\end{equation}
where $f$ represents mean squared error (MSE) in our study.


For training images with annotated label, both $L_{ArcFace}$ and $L_{Consist}$ are leveraged to optimize DeepNoise network, while for those unlabeled examples, only $L_{Consist}$ is adopted.

\textbf{Pseudo label softmax loss.} The pseudo label softmax loss $L_{pSoftmax}$ is introduced to minimize the difference between the averaged predictions derived from two teacher networks (in both ST-Xception and ST-Xception-Wide) and the output of each student network, which can be regarded as another constraint consistency loss function. 

\textbf{Model ensemble.} Model ensembling is an effective and commonly used approach in machine learning. Usually, it can efficaciously improve the overall performance of the network by aggregating predictions of each base model. It has been reported that model ensembling achieves satisfying performance on the segmentation of natural images \citep{liu2018path, zhao2019enhancing}, magnetic resonance images \citep{kamnitsas2017ensembles, li2018fully, chen2017hippocampus}, pathological images \citep{pimkin2018ensembling, qaiser2017tumor, zhao2019automated}, fundus images \citep{tang2019multi}, aerial images \citep{marmanis2016semantic}, and etc. In this work, we integrate two deep learning based models, i.e., ST-Xception and ST-Xception-Wide to derive the final classification results by averaging their respective predictions. Xception-Wide is developed based on Xception \citep{chollet2017xception} but with wider convolutional channels.

\textbf{Prior based post processing.} In R$\times$R$\times$1 dataset, same siRNA does not occur twice in an experimental plate. Based on this prior information, we add a post processing step to adjust the initial prediction of DeepNoise. The idea of the algorithm is as follows: For the popular classes (appear a few times more than minor classes), we decrease their predicted probabilities iteratively by a small value (an initial value is 0.001, and it is halved iteratively) until all classes are equally presented and the prediction score no longer improves. This balancing is not easy to enforce, but as a soft constraint in the loss, it works quite well.


\subsection*{Code availability} The source code of the proposed DeepNoise is available for research purposes at \url{https://github.com/Scu-sen/Recursion-Cellular-Image-Classification-Challenge}.

\section*{Acknowledgements}
We specifically express our gratitude to the Recursion organizer for their released dataset.

\bibliographystyle{unsrt}
\bibliography{nips}


\hfill 

\end{document}